\documentclass[12pt]{article}
\usepackage[T1]{fontenc} 
\usepackage{float}       
\usepackage{helvet}      
\usepackage{mathptmx}    
\usepackage{rsc}         
\usepackage{setspace}    
\usepackage{amsmath, amssymb, amsfonts}
\usepackage{graphicx,epstopdf}

\AtBeginDocument{\doublespacing}

\newfloat{chart}{htbp}{loc}
\newfloat{graph}{htbp}{loh}
\newfloat{scheme}{htbp}{los}

\usepackage[version=3]{mhchem} 

\makeatletter
\renewcommand{\@fnsymbol}[1]{\ensuremath{%
   \ifcase#1\or \mathsection\or *\or \dagger\or \ddagger\or
   \mathsection*\or \|\or \mathparagraph\or **\or \dagger\dagger
   \or \ddagger\ddagger \else\@ctrerr\fi}}
\makeatother

\begin{document}

\title{Fragility, Stokes-Einstein violation, \\
and correlated local excitations \\
in a coarse-grained model of an ionic liquid}

\author{
Daun Jeong,$^{1}$ 
M. Y. Choi,$^{2}$
Hyung J. Kim$^{3,4}$\thanks{Permanent address: Carnegie Mellon University}
\ and
YounJoon Jung$^{1}$\thanks{Corresponding author. E-mail: \texttt{yjjung@snu.ac.kr}}
}
\date{\small $^{1}$\textit{Department of Chemistry, Seoul National University, Seoul 151-747, Korea}\\ 
$^{2}$\textit{Department of Physics and Astronomy, Seoul National University, Seoul 151-747, Korea}\\
$^{3}$\textit{Department of Chemistry, Carnegie Mellon University, Pittsburgh, PA 15213, USA}\\
$^{4}$\textit{School of Computational Sciences, Korea Institute for Advanced Study, Seoul 130-722, Korea}\\
\today}

\maketitle

\begin{abstract} 
Dynamics of a coarse-grained model for the room-temperature ionic liquid, 1-ethyl-3-methylimidazolium hexafluorophosphate, couched in the united-atom site representation are studied via molecular dynamics simulations.  The dynamically heterogeneous behavior of the model resembles that of fragile supercooled liquids.  At or close to room temperature, the model ionic liquid exhibits slow dynamics, characterized by nonexponential structural relaxation and subdiffusive behavior.  The structural relaxation time,  closely related to the viscosity, shows a super-Arrhenius behavior.
Local excitations, defined as displacement of an ion exceeding a threshold distance, are found to be mainly responsible for structural relaxation in the alternating structure of cations and anions. As the temperature is lowered, excitations become progressively more correlated.  This results in the decoupling of exchange and persistence times, reflecting a violation of the Stokes-Einstein relation.
\end{abstract}

\section{Introduction}

Due to their important potential as environmentally benign alternatives to conventional toxic organic solvents,
room-temperature ionic liquids (RTILs) have attracted considerable attention recently.~\cite{seddon:rtil:CPP,keim:rtil,review2:rtil}
According to theoretical~\cite{shim:letter,*shim2,shim:rtil:rot,shim:rtil:review,margulis:rtil:dyn,bhargava:rtil,jeong:pol,jeong:1/f,kobrak:rtil:dyn1,*kobrak:rtil:dyn2,*kobrak:rtil:dyn3} and experimental~\cite{samanta:rtil1,*samanta:rtil2,*samanta:rtil3,*samanta:rtil4,maroncelli:rtil:solrot1,*maroncelli:rtil:complete1, *maroncelli:rtil:complete2,*maroncelli:rtil:solrot2, castner:il1,*castner:il2,vauthey:rtil:dyn,fayer:rtil:oke} studies on solvation and rotational dynamics of RTILs, their long-time behaviors are characterized by nonexponential decay.
This implies that RTILs are dynamically inhomogeneous and their local relaxation is widely distributed in time and space.~\cite{shim:rtil:rot,jeong:1/f} The clustered mobile and immobile ions observed in recent molecular dynamics (MD) simulation studies are ascribed to inhomogeneous dynamics in RTILs.~\cite{margulis:rtil:ree,Ngai:hetero}

Dynamic heterogeneity often invoked to explain many peculiar properties of  supercooled liquids~\cite{Richert:review:hetero,Ediger:review:hetero} refers to the enhanced temporal correlation of their local dynamic states with a decrease in temperature. From the viewpoint of facilitated motions, dynamics of supercooled liquids are dominated by fluctuations.~\cite{Chandler:prl,jung:exc} According to several studies based on lattice models, called kinetically-constrained models (KCMs),~\cite{FA,East} non-trivial structures in the space-time trajectory arising from dynamic constraints in the KCM description accurately reproduces  many of the dynamical properties of supercooled liquids.~\cite{Chandler:prl,Chandler:spacetime}   At the molecular level, it is found that trajectories of individual particles in atomistic models of supercooled liquids are in general governed by dynamic fluctuations and thus cannot be predicted from static properties, such as structures.~\cite{Berthier:predict} Meanwhile, recent studies have attempted to correlate length-scale dependent heterogeneous dynamics  with liquid structures on the basis of the dynamic propensity calculated from the isoconfigurational ensemble.~\cite{Harrowell:propensity1,*Harrowell:propensity2,*Harrowell:propensity3,*Daekeon:rtil:propen}  Despite these efforts, the origin of persisting dynamic correlations and the potential link of the dynamic correlations to structures still remain open questions in understanding dynamics of glassy liquids.

Glassy dynamics of supercooled liquids are characterized by many unique features such as the stretched exponential decay of time correlation functions, subdiffusive behavior in the intermediate time scale, decoupling of exchange and persistence times,~\cite{jung:exc} and breakdown of the Stokes-Einstein (SE) relation.~\cite{Kivelson:SE}  A variety of models of supercooled liquids, e.g., binary Lennard-Jones (LJ) mixture,~\cite{Chaudhuri:glass} supercooled water~\cite{Mazza:SE}, Weeks-Chandler-Anderson (WCA) mixture,~\cite{Hedges:decoupl} and KCMs,~\cite{jung:SE,jung:exc,Berthier:lifetime} reproduce these interesting features, which are generally believed to be intimately related to the dynamic heterogeneity.  This seemingly universal nature of dynamic heterogeneity and related phenomena is a main motivation to study glassy dynamics of ionic liquids,  the molecular details and interaction potentials of which are quite different from those of supercooled liquids. Specifically, strong electrostatic interactions in RTILs lead to an alternating structure of cations and anions, which is believed to generate  a significant memory effect on their dynamics.~\cite{jeong:1/f}  Since  the cage structure would exert a substantial influence on correlations between local excitation events, it would be both interesting and important to see if dynamic behaviors of RTILs bear any resemblance to those of supercooled liquids.

One of the  main challenges in studying dynamics of RTILs (and more generally glassy liquid systems) is the time scale.  As is well known, many glassy systems exhibit extremely slow dynamics, in particular, at low temperatures.   Therefore it is very difficult, if not impossible, to probe their long-time behaviors directly via atomistic MD simulations. However, it is precisely these dynamics at long times and their variations with temperatures that are of especial interest.  We thus employ a coarse-grained model to make simulations efficient and analyze long-time dynamics.  We note that this approach is not new.  Previous efforts\cite{jeong:thesis,voth:cg,*voth:am-cg,Chiappe:cg} based on similar descriptions have provided useful insights into  RTIL properties, both structure and dynamics.  In the present paper, we focus on the characterization of glassy dynamics and dynamic heterogeneity of RTILs.  As a prototypical RTIL, we study 1-ethyl-3-methylimidazolium hexafluorophosphate (${\rm EMI}^+{\rm PF_6}^-$).

To understand translational dynamics of EMI$^+$PF$_6^-$, we analyze the self-intermediate scattering function and mean square ion displacements of our model RTIL system.  We also examine the temperature dependence of its structural relaxation time and related fragility.\cite{Angell:fragility}  Our results suggest that EMI$^+$PF$_6^-$ is a fragile glass former and it violates the SE relation at low temperatures.  We regard correlations between local dynamic states as an essential aspect  of the aforementioned dynamic heterogeneity of glassy liquids.  Accordingly, we investigate distributions of the persistence and exchange times, i.e., waiting times for the first and subsequent excitations, by monitoring the motions of individual ions.  To understand the influence of Coulombic interactions on glassy dynamics, we also make contact with non-ionic model systems of supercooled liquids. 

The outline of this paper is as follows: In Sec.~\ref{sec:model}, we introduce the coarse-grained model of an ionic liquid and briefly describe the simulation methods. In Sec.~\ref{sec:results}, we present the MD results on glassy dynamics, including fragility, nonexponential relaxation, subdiffusion, and violation of the SE relation.  The correlation of local events together with the decoupling of persistence and exchange times is demonstrated  in Sec.~\ref{sec:anal}, while the roles of the Coulombic interactions are discussed in Sec.~\ref{sec:coul}.  Concluding remarks are offered in Sec.~\ref{sec:conclusion}.

\section{RTIL models}

\label{sec:model}
\begin{figure}
\centering
\resizebox{\textwidth}{!}{
		\includegraphics{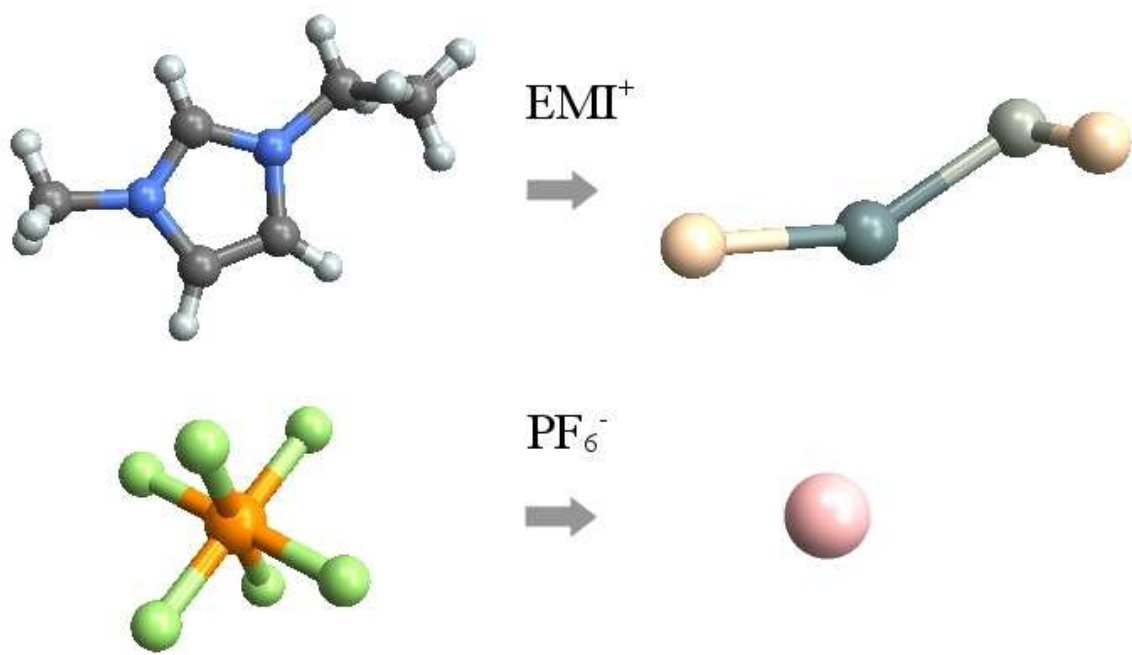}
}
\caption{Coarse-graining scheme.  ${\rm EMI}^+$ is reduced to a 4-atom cation, where 5 atoms of the imidazolium ring of ${\rm EMI}^+$ and 3 hydrogen atoms directly attached to it are collapsed to a single united atom (T1) in the coarse-grained model. The methyl and ethyl groups of ${\rm EMI}^+$ are represented, respectively, as one-atom (M1) and two-atom (M3 and E1) moieties and ${\rm PF}_6^-$ is simplified as a united atom as in Ref.~\citenum{shim1}.}
\label{fig:cg}
\end{figure}

In this section, we give a brief explanation of our coarse-grained model\cite{jeong:thesis} for ${\rm EMI}^+{\rm PF_6}^-$ 
and compare its results for structure and translational dynamics with those of a more atomistic description. 


\subsection{Coarse-grained model}
\label{sec:model:cgm}

Our coarse-grained model is based on the RTIL description used by Kim and coworkers\cite{shim1} to study various processes in RTILs.\cite{shim:letter,*shim2,shim:rtil:rot,jeong:1/f,shim:rtil:et,*shim:rtil:sn1,*shim:rtil:diex}  Specifically, they employed the united atom representation for the methyl group (M1) as well as the CH$_2$ (E1) and CH$_3$ (M3) moieties of the ethyl group of cations.  They used the AMBER force field\cite{amber} for the Lennard-Jones (LJ) interactions and partial charge assignments\cite{lynden-bell:parameter} proposed by Lynden-Bell and coworkers for Coulombic interactions.   ${\rm PF_6}^-$ was also described as a united atom.  Hereafter, this model will be referred to as the AM description.  In our coarse-grained model (CGM), we further simplify the cation description by representing the imidazole ring and H atoms directly attached to it as a single atomic site T1 (Fig.~\ref{fig:cg}).  Thus each EMI$^+$ ion consists of 4 united-atom sites, M1, M3, E1 and T1 in CGM.   The LJ parameters of T1 were adjusted, so that CGM reproduces the liquid structure of the atomistic AM description reasonably well (see below).  For M1, M3,  E1 and PF$_6^-$, we used the parameters of AM without any further adjustment.  

\begin{table}[!t]
\centering
\caption{The LJ parameters, partial charges, and masses of coarse-grained atoms}
\vspace*{15pt}
\begin{tabular}{c|cccc}
\hline
atom & $\sigma_{ii}$ (\AA)  & $\epsilon_{ii}$ (kJ/mol) & $q_i$ $(e)$ & mass (amu) \\
\hline
M1  &  3.905 & 0.7330 &    0.316   & 15.04092 \\
T1  &  4.800 & 1.5000 &    0.368   & 67.08860 \\
E1  &  3.800 & 0.4943 &    0.240   & 14.03298 \\
M3  &  3.800 & 0.7540 &    0.076   & 15.04092 \\
PF6 &  5.600 & 1.6680 &   -1.000   & 144.97440 \\
\hline
\end{tabular}

\label{table:param}
\end{table}

We performed MD simulations of EMI$^+$PF$_6^-$ in both the AM and CGM representations using the DL\_POLY program.\cite{dlpoly}. Atoms $i$ and $j$ at positions ${\bf r}_i$ and ${\bf r}_j$ interact with each other through the LJ and Coulomb potentials: \begin{equation}
U_{ij}=4\epsilon_{ij}\left[ \left(\frac{\sigma_{ij}}{r_{ij}}\right)^{12}-
\left(\frac{\sigma_{ij}}{r_{ij}}\right)^{6} \right]+\frac{q_iq_j}{r_{ij}},
\end{equation}
where $r_{ij}\equiv |{\bf r}_i -{\bf r}_j|$ is the distance between the two atoms. The parameters of CGM employed in the present study are compiled  in Table~\ref{table:param}.  For the AM parameters, the reader is referred to Ref.~\citenum{shim1}.  For LJ interactions between unlike atoms, the Lorentz-Berthelot combining rules were used.~\cite{Allen}

The simulation cell of the CGM ionic liquid comprises 512 pairs of rigid cations and anions. We performed simulations in the canonical ensemble at six different temperatures, $T=300, 350, 400, 475, 600$ and $800\,{\rm K}$, using the Nos{\'e}-Hoover thermostat and at density $\rho = 1.31\,{\rm g/cm^{3}}$.  Periodic and cubic boundary conditions were employed and long-range electrostatic interactions were computed via the Ewald method.  Starting from a crystal configuration obtained by alternating the cations and anions, we equilibrated the system for $2\,{\rm ns}$ prior to production runs at $800\,{\rm K}$.  At lower temperatures, we used as an initial configuration one of the equilibrated configurations at higher $T$ that is closest to the temperature of the system under consideration and equilibrated the system for $2\,{\rm ns}$ at $600\,{\rm K}$ and $475\,{\rm K}$, $5\,{\rm ns}$ at $400\,{\rm K}$, and $10\,{\rm ns}$ at $350\,{\rm K}$ and $300\,{\rm K}$.  Production runs following equilibration were $5\,{\rm ns}$ in length for $800\,{\rm K}$ and $600\,{\rm K}$, $10\,{\rm ns}$ for $475\,{\rm K}$ and $400\,{\rm K}$, $50\,{\rm ns}$ for $350\,{\rm K}$, and $60\,{\rm ns}$ for $300\,{\rm K}$.  At each thermodynamic condition, we carried out six independent production runs, from which the averages were computed.  Thus we used, for example, six $60\,{\rm ns}$ trajectories to analyze various dynamic quantities at $300\,{\rm K}$ in CGM.

In the case of AM, we considered 112 pairs of EMI$^+$ and PF$_6^-$ ions with $\rho = 1.31\,{\rm g/cm^{3}}$ at $T=350$ and $400\,{\rm K}$.   At either temperature, we performed three independent simulations, each of which was carried out with $10\,{\rm ns}$ equilibration, followed by a $40\,{\rm ns}$ trajectory.  Ensemble averages were calculated using three trajectories thus obtained.  

\begin{figure}
\centering
\resizebox{\textwidth}{!}{
	\includegraphics{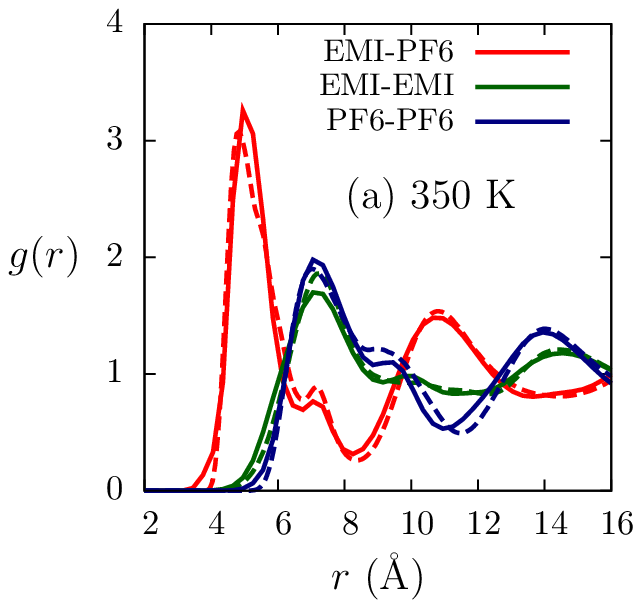}
	\includegraphics{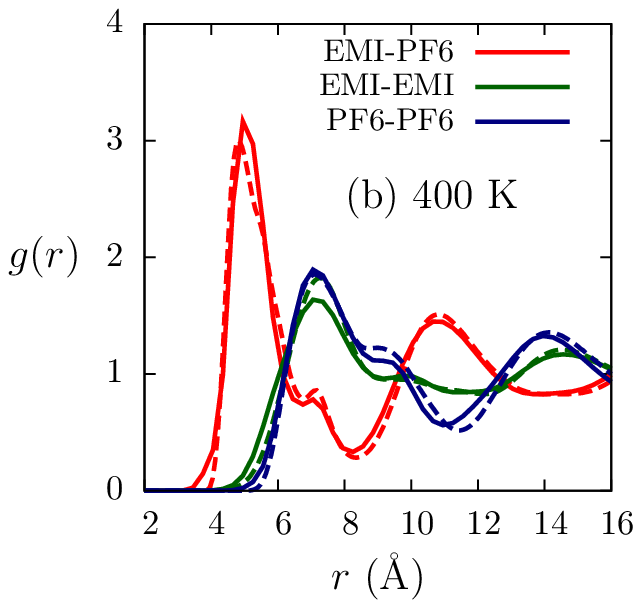} 
}
\caption{Radial distributions of ions in EMI$^+$PF$_6^-$ at (a) $350\,{\rm K}$ and (b) $400\,{\rm K}$. The results of CGM and AM are plotted in solid and dashed lines, respectively.  The center of mass position is used to represent cation locations in CGM and AM.}
\label{fig:str}
\end{figure}

\subsection{Structure}
\label{sec:model:structure}

Here we consider MD results of the CGM and AM for structure to gain insight into how realistic the former description is. In Fig.~\ref{fig:str}, their results for radial distributions of ions at $T = 350\,{\rm K}$ and $400\,{\rm K}$ are compared.  There we employed  the center of mass to represent the cation positions for CGM and AM.  We notice that CGM captures the RTIL structure of AM very well. The two models yield a excellent  agreement in both the peak positions and heights, including minor secondary peaks, e.g., structure near $7\,{\textrm \AA}$ in the cation-anion distribution.  Even in the case of the main peak of the cation-anion distribution which shows the largest deviation between the two, the discrepancy in peak location is only $\sim\!0.02\,{\textrm \AA}$.  Considering the drastic approximation of the planar imidazole ring as a spherical atom, we think that overall the coarse-grained model fares very well in reproducing the RTIL structure of AM.


\subsection{Translational dynamics}\label{sec:model:translation}
To understand the effect of our coarse-graining (Fig.~\ref{fig:cg}) on system dynamics, we examine ion translational motions by considering their mean square displacement, 
$\Delta (t)=\langle N^{-1}\sum_{i=1}^{N}|{\bf r}_i(t)-{\bf r}_i(0)|^2 \rangle$,
and self-intermediate scattering function, 
$F_s(q_0,t)\equiv \langle N^{-1} \sum_{i=1}^{N} e^{i {\bf q}_0\cdot[{\bf r}_i(t)-{\bf r}_i(0)]}\rangle$. 
Here $\langle \cdots \rangle$ denotes the equilibrium ensemble average, $N$  the number of ions, and ${\bf q}_0$ the wave vector corresponding to the position of the first peak in the static structure factor.

\begin{figure}
\centering
\resizebox{\textwidth}{!}{
	\includegraphics{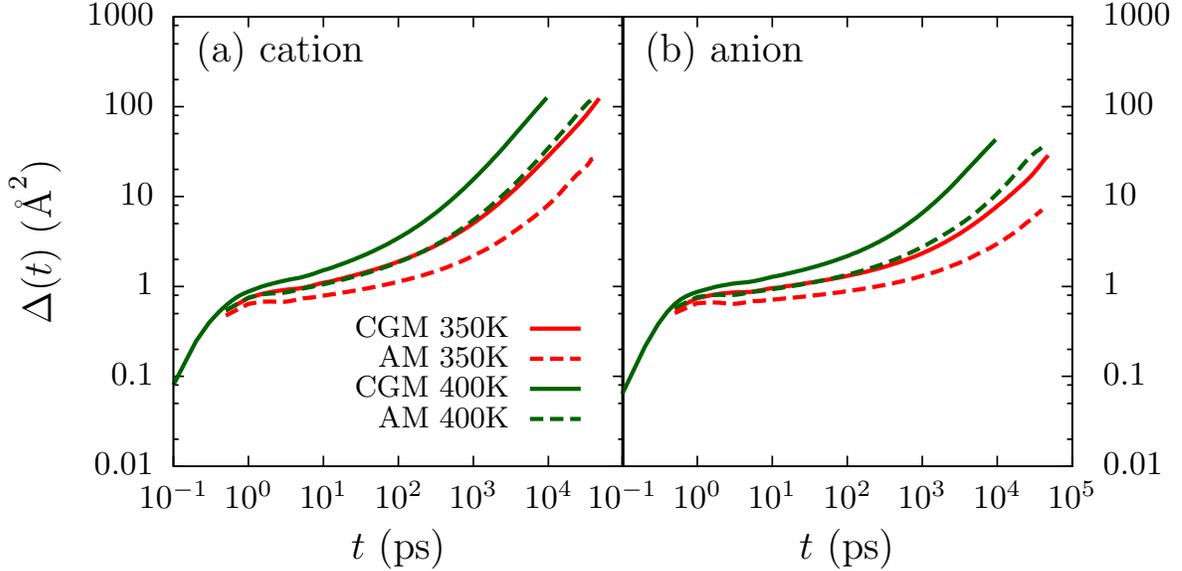}
}
\caption{Mean square displacement $\Delta(t)$ of (a) cations and (b) anions in CGM and AM at $T=350\,{\rm K}$ and $400\,{\rm K}$.}
\label{fig:msd_c}
\end{figure}
\begin{figure}
\centering
\resizebox{\textwidth}{!}{
	\includegraphics{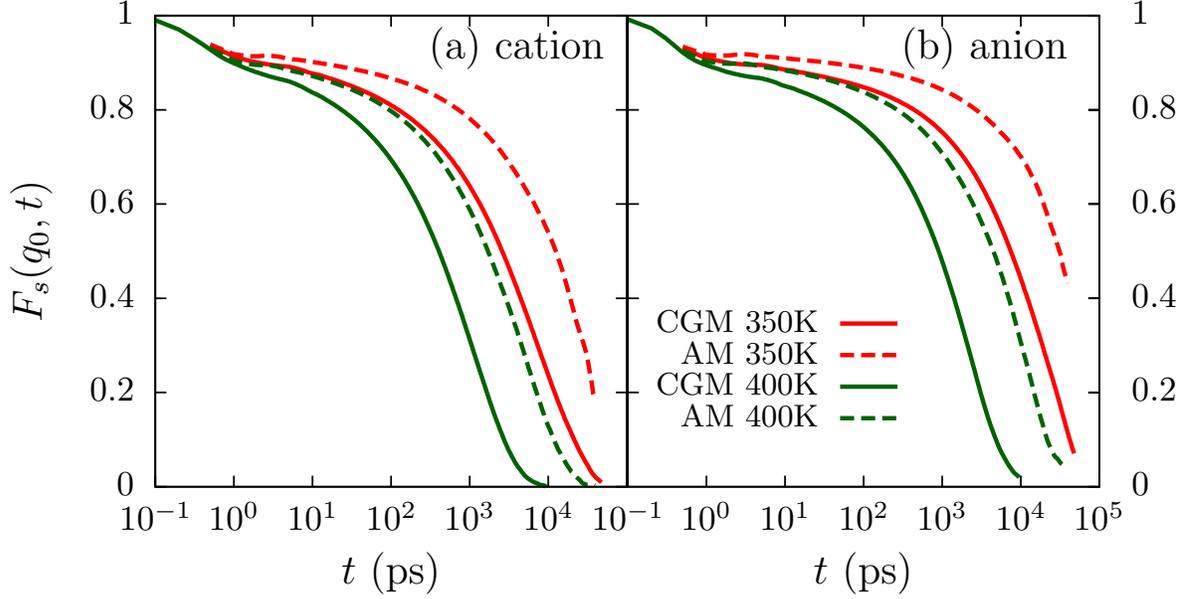}
}
\caption{Self-intermediate scattering function $F_s(q_0,t)$ for (a) cations and (b) anions in CGM and AM at $T=350\,{\rm  K}$ and $400\,{\rm K}$.  The wave vector employed in (a) and (b) are, respectively, $q_0=0.858\,\textrm{\AA}^{-1}$ and $0.878 \,\textrm{\AA}^{-1}$, corresponding to the positions of the first peak in their static structure factor. 
}
\label{fig:isf_c}
\end{figure}
\begin{table}[!t]
\centering
\caption{MD results for the diffusion constant $D$ and the structural relaxation time $\tau_{\alpha}$ for CGM}
\vspace*{15pt}
\begin{tabular}{c cc cc}
\hline
$T$ (K) & \multicolumn{2}{c}{$D$ ($\textrm{\AA}^2/{\rm ps}$) } & 
\multicolumn{2}{c}{$\tau_{\alpha}$ (ps)} \\
 & cation  & anion & cation & anion \\
\hline
\noalign{\vspace*{5pt}}
300  &  $5.4\times 10^{-5}$& $9.6\times 10^{-6}$& 59600  & 244000 \\   
350  & $4.3\times 10^{-4}$ &  $9.7\times 10^{-5}$& 5030    & 12700     \\     
400  & $2.2\times 10^{-3}$ &  $7.2\times 10^{-4}$& 779      &  1610      \\     
475  & $8.8\times 10^{-3}$ &  $4.0\times 10^{-3}$&  154       &  261        \\      
600  & $3.3\times 10^{-2}$ &  $2.0\times 10^{-2}$ & 36.1      &  50.2       \\       
800  & $9.9\times 10^{-2}$ &  $6.8\times 10^{-2}$  & 11.4      &  13.5         \\
\noalign{\vspace*{5pt}}
\hline
\end{tabular}
\label{table:MDresults}
\end{table}

MD results in Figs. \ref{fig:msd_c} and~\ref{fig:isf_c} show that the coarse-grained model exhibits subdiffusive behavior and nonexponential relaxation, consistent with the atomistic model.  We, however, notice that dynamics of the former proceed faster than the latter. To quantify this, we calculated the diffusion constant 
$D=\lim_{t\rightarrow \infty} [6(t-t_0)]^{-1}{\langle N^{-1}\sum_{i=1}^{N}[{\bf r}_i(t)-{\bf r}_i(t_0)]^2\rangle}$,  
where $t_0$ denotes the time when the Fickian behaviors appear in $\Delta (t)$,
and determined the structural relaxation time $\tau_{\alpha}$ via
\begin{equation}
\label{eq:taualpha}
F_s(q_0,\tau_{\alpha})={\rm e}^{-1}\ .
\end{equation} 
The MD results for $D$ and $\tau_{\alpha}$ are compiled in Table~\ref{table:MDresults}. At $400\,{\rm K}$, we obtained $D=2.2 \times 10^{-3}\, {\textrm \AA}^2$/ps and 7.2 $\times 10^{-4}\, {\textrm \AA}^2$/ps for cations and anions,  respectively, for CGM, while the corresponding values for AM were 6.5 $\times 10^{-4}\, {\textrm \AA}^2$/ps and 1.4 $\times 10^{-4}\, {\textrm \AA}^2$/ps. As for $\tau_{\alpha}$, the coarse-grained model yields 779\,ps and 1610\,ps for cations and anions, respectively, whereas the AM description results in significantly longer relaxation times, 3680\,ps and 9010\,ps.  Because of this discrepancy, i.e., CGM is faster than AM in dynamics, $\Delta(t)$ and $F_s(q,t)$ of AM at $400\,{\rm K}$ are in better accord with those of CGM at $350\,{\rm K}$ than at $400\,{\rm K}$ (Figs. \ref{fig:msd_c} and~\ref{fig:isf_c}). (For CGM at $350\,{\rm K}$, simulations yield $D=4.3\times10^{-4}{\textrm \AA}^2$/ps and $9.7\times10^{-5} {\textrm \AA}^2$/ps, and $\tau_{\alpha}=5030$\,ps and $12700 \,{\rm ps}$ for cations and anions, respectively.) This is not surprising in that structure and dynamics of liquid systems, including ionic liquids, depend on molecular shape (i.e., LJ interactions) and charge distributions (viz., electrostatic interaction) of constituent particles.~\cite{Chiappe:cg,voth:cg}  In addition to the difference in cation geometry, negative partial charges of nitrogen atoms of EMI$^+$ present in the AM description are completely absent in CGM because of the united atom representation of the imidazole ring.  This simplification reduces charge anisotropy of cations and thus frustration in the structure and dynamics of CGM. We therefore expect that both rotational and translational dynamics would be accelerated in CGM, compared to AM. Here we consider only the translational dynamics of cations and anions and postpone the rotational dynamics for a future study.
  
\section{Glassy dynamics}
\label{sec:glassy}
In this section, we analyze the glassy behavior of the coarse-grained RTIL with the aid of MD simulation results. We examine structural relaxation and ion diffusion at various temperatures and demonstrate that our model belongs to the class of fragile glass formers and violates the Stokes-Einstein relation.
\label{sec:results}

\subsection{Structural relaxation and fragility}
\label{sec:fragility}
\begin{figure}
\centering
\resizebox{\textwidth}{!}{
	\includegraphics{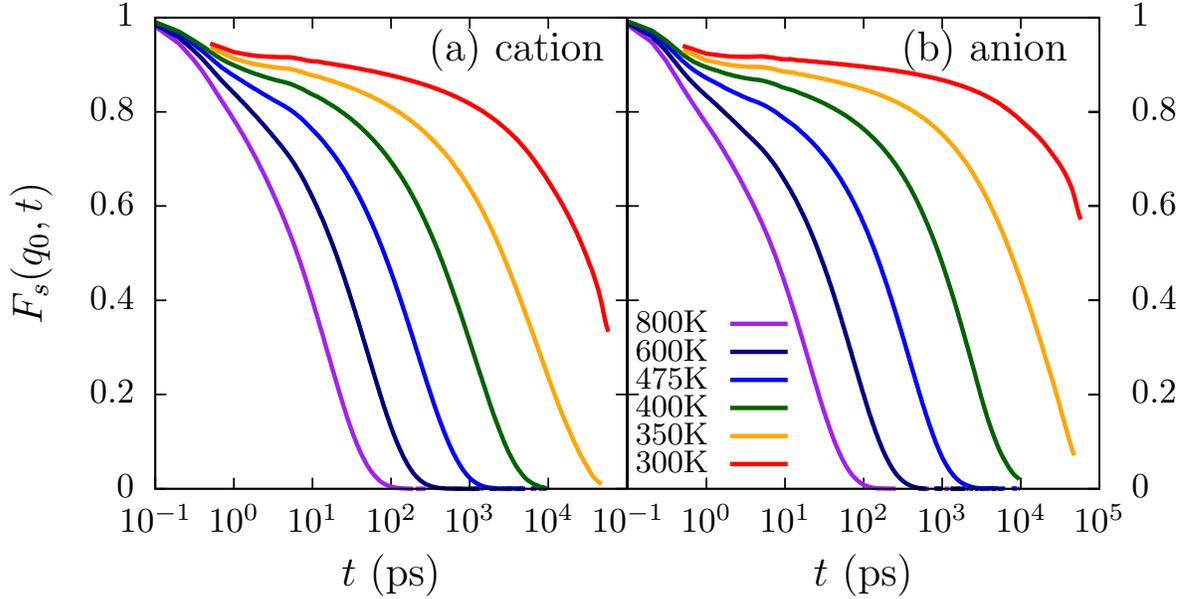}
}
\caption{Self-intermediate scattering function $F_s(q_0,t)$ for (a) cations and (b) anions   in a coarse-grained ionic liquid at various temperatures. The wave vector $q_0$ is the same as in Fig.~\ref{fig:isf_c}. 
In the $\alpha$-relaxation regime, $F_s(q_0,t)$ exhibits nonexponential decay, which is well described by $c\exp[-(t/\tau_0)^\beta]$ for all temperatures. }
\label{fig:ISF}
\end{figure}

The structural relaxation through ion translational dynamics is usually described by the self-intermediate scattering function $F_s(q_0,t)$.  In Fig.~\ref{fig:ISF}, the CGM predictions for $F_s(q_0,t)$ at six different temperatures are displayed.  At high $T$, the ionic liquid  behaves almost like a normal liquid; the thermal fluctuations dominate over the constraints of caged structures of ions, mainly formed by counterions.  As the temperature decreases, characteristics of glassy dynamics, such as subdiffusivity  and nonexponential relaxation, become more apparent.  

At low $T$, the presence of the plateau regime ($\beta$ relaxation) and slow $\alpha$ relaxation, which are hallmarks of supercooled liquids, is rather prominent.  The contribution of inertial dynamics  in the first $0.1\,{\rm ps}$ or so accounts for less than 10\% of the entire relaxation of $F_s(q_0,t)$, and structural correlation persists for several decades in time, thereby indicating the highly viscous RTIL environment. The slow nonexponential relaxation subsequent to the plateau regime is well described by a stretched exponential function, $c \exp[-(t/\tau_0)^{\beta}]$.  At $300\,{\rm K}$, the exponent $\beta$ is found to be 0.64 and 0.59 for cations and anions, respectively.  A substantial deviation of these $\beta$ values from unity is another good indicator of the glassy dynamics in the RTIL system.  As $T$ increases, so does $\beta$.  At $800\,{\rm K}$, the highest temperature we studied, the $\beta$ values are 0.89 and 0.92.  Even though greatly enhanced thermal fluctuations at this temperature accelerate structural relaxation immensely by more than four orders of magnitude compared to that of $300\,{\rm K}$, $F_s(q_0,t)$ still maintains a nonexponential character, presumably due to its high pressure condition.  We thus expect that if we lower the pressure by reducing its density, the structural relaxation would become a single-exponential decay.

\begin{figure}
\centering
\resizebox{0.95\textwidth}{!}{
	\includegraphics{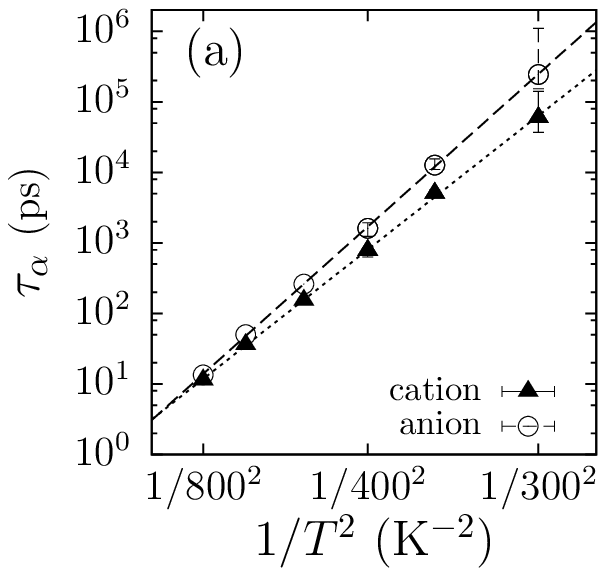}
	\includegraphics{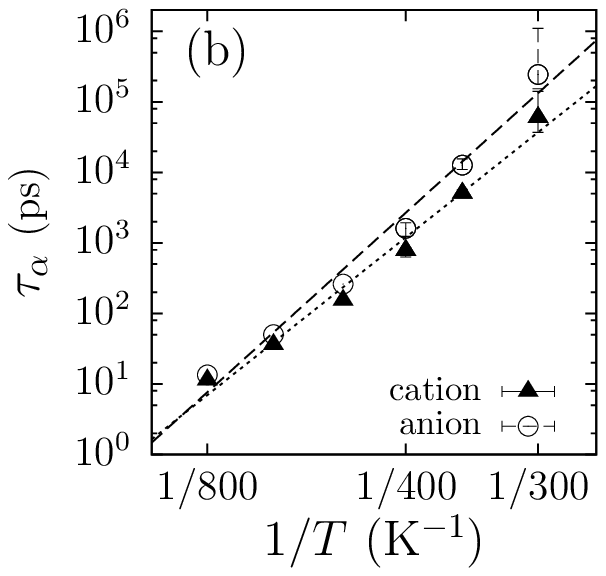}
}
\caption{Temperature dependence of structural relaxation time $\tau_\alpha$, Eq.~\ref{eq:taualpha}. Dashed and dotted lines are the fits to MD results using (a) $\tau_\alpha=c_1 \exp(d_1/T^2)$ and (b) $\tau_\alpha=c_2 \exp(d_2/T)$. Error bars represent the maximum and the minimum values of $\tau_\alpha$ among six independent  trajectories.}
\label{fig:SA}
\end{figure}

We turn to the temperature dependence of $\tau_\alpha$ in the CGM description presented in Figure~\ref{fig:SA} and Table~\ref{table:MDresults}.  Since slow structural relaxation at $300\,{\rm K}$ does not allow the determination of its $\tau_\alpha$ directly from the simulation results, we estimated it by employing a stretched exponential fit to $F_s(q_0,t)$.  At all other temperatures, $\tau_\alpha$ was determined using the MD results for $F_s(q_0,t)$ in Eq.~\ref{eq:taualpha}. The most salient aspect of our results  in Figure~\ref{fig:SA} is that $\tau_{\alpha}$ does not follow the Arrhenius law $\tau_{\alpha}\propto\exp(d_2/T)$.  Rather the structural relaxation time shows a super-Arrhenius behavior; specifically, it varies with the temperature as $\tau_{\alpha}\propto\exp(d_1/T^2)$.   This means that the CGM RTIL studied here resembles a fragile glass former in the temperature dependence.

For clarity, we make a couple of remarks here. First, while the RTIL density is assumed to be fixed in the present study, it tends to decrease with increasing temperature for real ionic liquids.  This density variation, if incorporated, would lead to acceleration of structural relaxation in CGM at high $T$, compared to the results in Fig.~\ref{fig:SA}.   This would in turn strengthen the super-Arrhenius character of $\tau_\alpha$ and thus enhance the fragile behavior of CGM. Second, as pointed out in Sec.~\ref{sec:model:translation} above, temperatures of CGM and AM do not coincide.  In other words, the temperature of CGM does not correspond to the actual temperature of the atomistic system.  As a consequence, our finding that EMI$^+$PF$_6^-$ is a fragile glass former based on the CGM description might not be directly applicable to the real ionic liquid.  We however note that  a super-Arrhenius temperature dependence was observed in recent measurements in a similar RTIL.\cite{richert:fragile} We thus believe that our result on the fragility of EMI$^+$PF$_6^-$ is robust.

\subsection{Breakdown of the Stokes-Einstein Relation}
\label{sec:SE}
In a normal liquid, the diffusion constant $D$ is usually related to the viscosity $\eta$ via the Stokes-Einstein relation 
\begin{equation}
D \propto \frac{T}{\eta}\ .
\label{eq:SE}
\end{equation}
For convenience, we consider another relation
\begin{equation}
D \propto \frac{1}{\tau_\alpha}\ ,
\label{eq:SE2}
\end{equation}
which is equivalent to Eq.~\ref{eq:SE} if the structural relaxation time is proportional to $\eta/T$.  Eq.~\ref{eq:SE2} results when translational motions of constituent particles are described by a normal diffusion equation, the Gaussian solution of which is $F_s(q_0,t)=\exp (-q_0^2 Dt) \equiv \exp (-t/\tau_\alpha)$.


\begin{figure}
\centering
\resizebox{\textwidth}{!}{
	\includegraphics{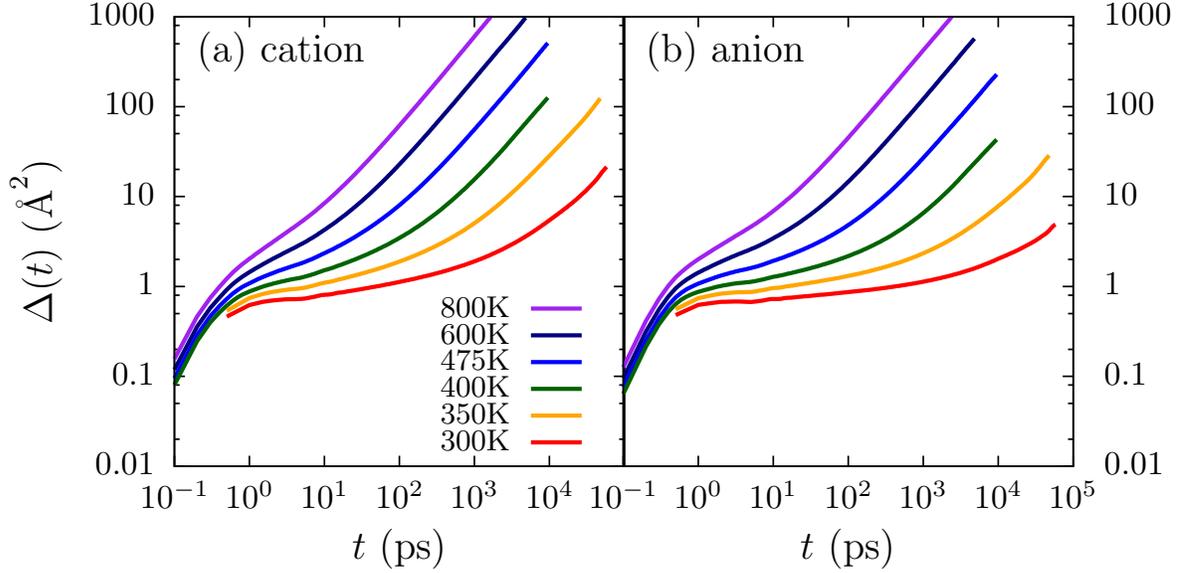}
}
\caption{Mean square displacement of (a) cations and (b) anions in the CGM description.  }
\label{fig:MSD}
\end{figure}

Simulation results for the mean square displacement of cations and anions at different temperatures are displayed in Fig.~\ref{fig:MSD}.  Subdiffusion in the intermediate time scale, another common feature of supercooled liquids, is quite pronounced, especially at low $T$.   At long times, ion translations tend to become normal diffusion. We notice that
the time scale associated with the transition from non-Fickian to Fickian dynamics generally coincides with the structural relaxation time.~\cite{szamel:Fick_time}  Another noteworthy feature is that
for both $\Delta(t)$ and $F_s(q_0,t)$, motions of PF$_6^-$ tend to be slower than those of EMI$^+$.  This is attributed to the fact that anions are heavier than cations. 

\begin{figure}
\centering
\resizebox{\textwidth}{!}{
	\includegraphics{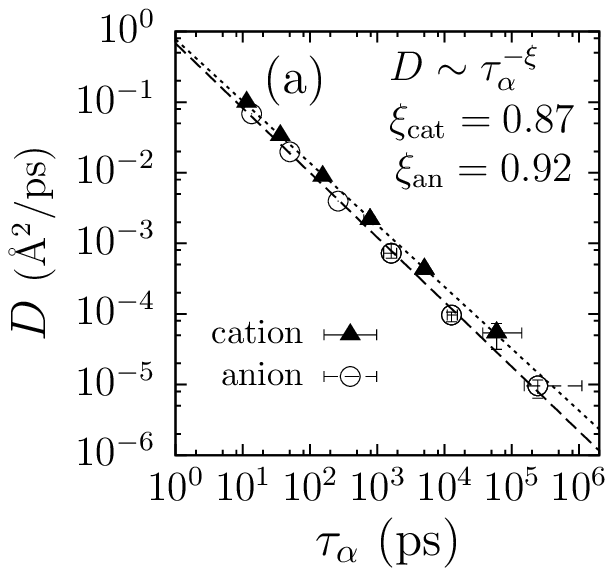}
	\includegraphics{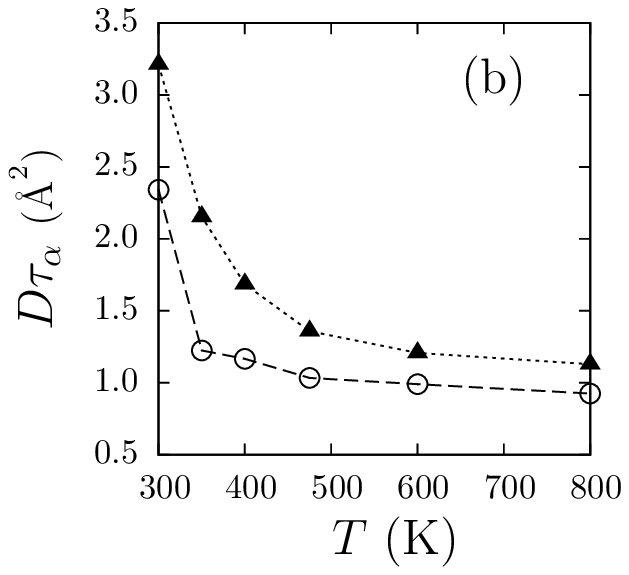}
}
\caption{Breakdown of the SE relation in the coarse-grained ionic liquid. (a) Dashed and dotted lines are the fits with the scaling relation $D\sim\tau_\alpha^{-\xi}$, where the exponent $\xi$ is given by 0.87 and 0.92 for cations and anions, respectively. (b) $D\tau_{\alpha}$ deviates from 
the constant value as the temperature decreases.  Lines are drawn merely as a guide for the eyes.}
\label{fig:SE}
\end{figure}

For additional insight, we have analyzed the relation between $D$ and $\tau_\alpha$ via
\begin{equation}
\label{eq:Dxi}
D\sim\tau_\alpha ^{-\xi},
\end{equation} 
where $\xi=1$ corresponds to the SE relation.
We found that $\xi=0.87$ and $0.92$ for cations and anions, respectively (Fig.~\ref{fig:SE}(a)).  Thus the product $D\tau_\alpha$ develops a positive deviation from a constant value as $T$ decreases (Fig.~\ref{fig:SE}(b)).
This reveals a  weak violation of the SE relation for our model RTIL system.

The breakdown of the SE relation indicates that translational dynamics of the ionic liquid can not be described by the conventional diffusion equation, which is a continuity equation combined with a constitutive relation given by Fick's law.  Specifically, it is assumed that the diffusion current is proportional to the spatial gradient of the particle concentration and obtains stationarity instantaneously in response to the external perturbation. Accordingly, the diffusion equation is valid only in the limit where the time is sufficiently coarse-grained to ensure instantaneous establishment of stationarity. If there is a significant delay in time before the system reaches stationarity,  a crossover from the non-Fickian regime, characterized by anomalous diffusion and $\beta$ relaxation, to the Fickian regime happens. We attribute the main cause of the delay to cage dynamics, i.e., immobile cages that last for a long time (see below). 
In this sense, the violation of the SE  relation is a manifestation of dynamic heterogeneity,~\cite{jung:SE} which we turn to next.

\section{Correlated local excitations}
\label{sec:anal}
\subsection{Decoupling of exchange and persistence times}
\label{sec:decoupl}
\begin{figure}
\centering
\resizebox{\textwidth}{!}{
	\includegraphics{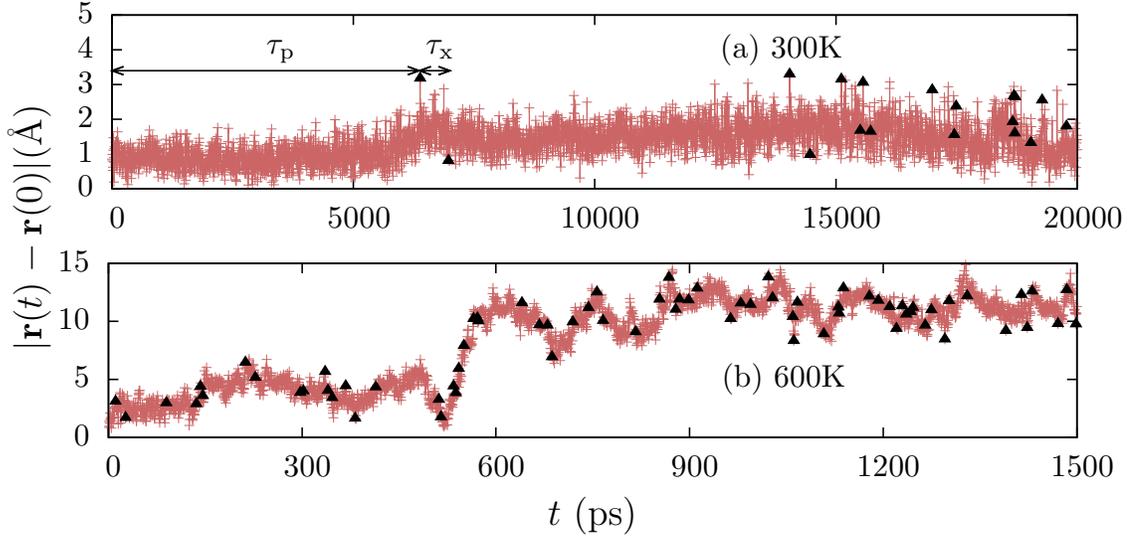}
}
\caption{Displacement of a cation from simulation trajectories at (a) $300\,{\rm K}$ and (b) $600\,{\rm K}$. 
Occurrences of excitations at which the cation moves beyond $d$(=3.0 \AA) are marked with triangles. For a given initial position, the waiting time until the occurrence of the first excitation defines the persistence time $\tau_{\mathrm p}$, while the time 
interval between subsequent excitations yields the exchange times $\tau_{\mathrm x}$.
}
\label{fig:traj}
\end{figure}

In ionic liquids, due to electrostatic interactions, a central ion is surrounded mainly by counterions in its immediate neighborhood, termed first solvation shell.  Thus dynamics of the central ion will be influenced by those of the first solvation shell (cage) and vice versa.  (In normal electrolytes, for example, this counter-ion cage, referred to as an ion atmosphere, tends to reduce the diffusion constant of the central ion at low concentration.) Suppose that the central ion undergoes thermal motions in the cage.  Once in a while a large fluctuation enables it to escape the cage and subsequently the cage reorganizes. As the temperature is lowered, thermal fluctuations become suppressed and as a result, the frequency of the ion escape from the cage diminishes.  While this picture may need quantitative elaboration, it is nonetheless useful to obtain a qualitative understanding of ``excitations'' introduced below and their properties in glassy environments.

We first introduce an excitation as a local event that an ion moves over a distance exceeding a threshold value $d$. The persistence and exchange times associated with excitations are then defined as follows:~\cite{jung:exc,Hedges:decoupl} the persistence time $\tau_{\rm p}$ is the waiting time $t_1$ for an ion $i$ to undergo its first excitation such that $|{\bf r}_i(t_1)-{\bf r}_i(0)|\geq d$, and the exchange time $\tau_{\rm x}$ includes a set of time intervals ${t_2, t_3,\cdots}$ between subsequent excitation events, i.e., $|{\bf r}_i(t_1+t_2)-{\bf r}_i(t_1)|\geq d$, $|{\bf r}_i(t_1+t_2+t_3)-{\bf r}_i(t_1+t_2)|\geq d$, etc. Accordingly, the frequency of excitations gauges the mobility of ions.  In the present study, we employ 3.0~\AA\ for $d$.  We will return to discuss other possibilities for $d$ later on.

In a glassy environment where cage reorganization is slow, the likelihood of an ion undergoing a second excitation after its first one is higher than that of the initial excitation because the liquid structure disturbed by the first excitation generally provides a favorable environment for another excitation. 
In other words, the excitations are not governed by a Poisson process.
In Fig.~\ref{fig:traj}, typical trajectories of an ion at $300\,{\rm K}$ and $600\,{\rm K}$ are displayed.  On each trajectory, the events of local excitations are marked with triangular symbols. As expected, excitations at $300\,{\rm K}$ are rare events; their occurrences are irregular and intermittent.  This, for instance, leads to a non-exponential tail on the long-time end of the distribution of  exchange times, i.e., time intervals between two consecutive excitations (see below).  By contrast, the trajectory at $600\,{\rm K}$ is characterized by more regular and frequent excitations than that at 300\,K.  

\begin{figure}
\resizebox{\textwidth}{!}{
	\centering
	\includegraphics{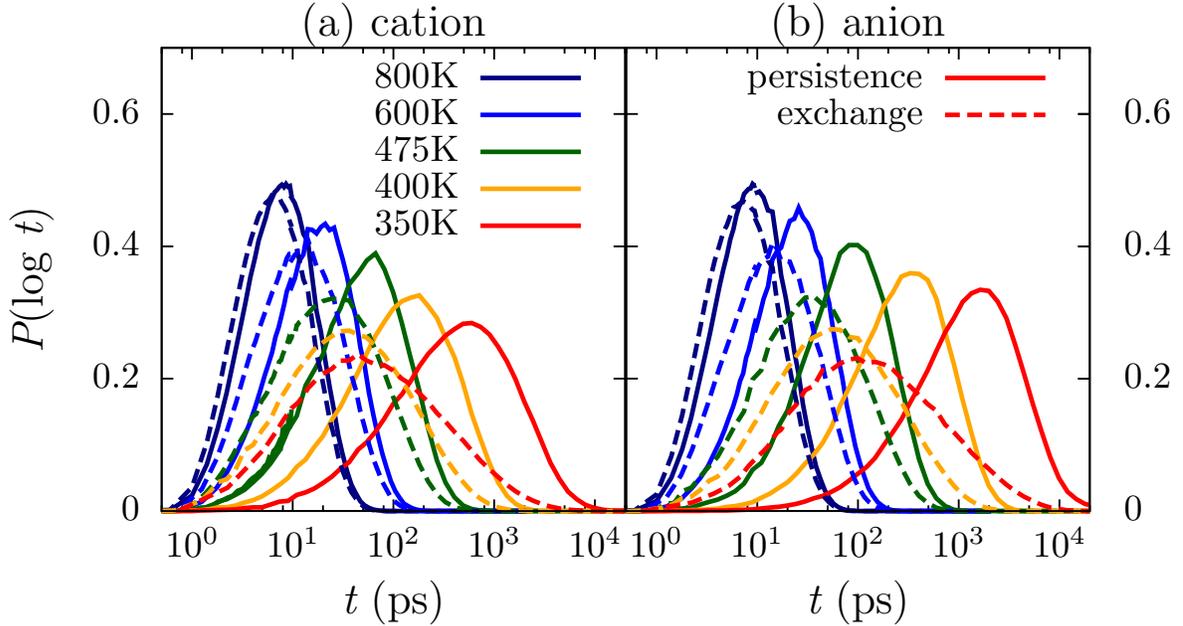}
}
\caption{Decoupling of persistence and exchange times for (a) cations and (b) anions in a coarse-grained ionic liquid. 
The persistence time $\tau_{\mathrm p}$ and the exchange time $\tau_{\mathrm x}$ are defined as the waiting times for an ion to produce a displacement exceeding $d$, taken as $3.0\,{\textrm \AA}$,  for the first time and thereafter, respectively. 
The solid and dashed lines correspond to the probability distributions of $\log \tau_{\mathrm p}$ and $\log \tau_{\mathrm x}$, respectively.}
\label{fig:decoupl}
\end{figure}

\begin{figure}
\centering	
\resizebox{0.5\textwidth}{!}{
	\includegraphics{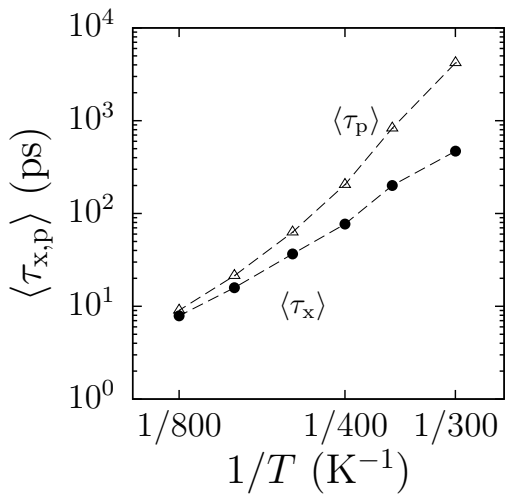}
}
\caption{Average persistence and exchange times versus the inverse temperature for $d= 3.0\,\textrm{\AA}$. Lines are drawn as a guide for the eyes.}
\label{fig:txtp}
\end{figure}

Figure~\ref{fig:decoupl} exhibits the probability distributions of the logarithms of persistence and exchange times for cations and anions. At $800\,{\rm K}$, the persistence and exchange times show nearly identical distributions.  This clearly indicates that the excitation events follow the Poisson process, viz., they are not correlated. As $T$ decreases, two distributions become distinct and their difference increases.  The center of the distribution for $\tau_{\rm p}$ shifts to longer time much more rapidly than the corresponding distribution for $\tau_{\rm x}$.   As a result, the deviation between the average exchange time $\langle\tau_{\rm x}\rangle$ and average persistence time $\langle\tau_{\rm p}\rangle$ rises markedly with lowering $T$ (see Fig.~\ref{fig:txtp}). This decoupling of $\tau_{\rm x}$ and $\tau_{\rm p}$ observed here shows that the excitations become increasingly more correlated as $T$ decreases, exposing the dynamically heterogeneous nature of our RTIL system at low $T$. We also notice dramatic enhancement of $\langle\tau_{\rm p}\rangle$ at low $T$, compared to $\langle\tau_{\rm x}\rangle$.   For instance, $\langle\tau_{\rm p}\rangle$ is longer than $\langle\tau_{\rm x}\rangle$ by one order of magnitude at 300\,K. This suggests the development of persisting immobile regions. According to a recent MD study in a similar RTIL,\cite{Zhao:ionpair} the life time of stable contact ion pairs that seldom move a large distance as a pair can exceed a few nanoseconds. Such long-lived pairs could potentially be a candidate for immobile regions associated with long $\tau_{\rm p}$. Finally,  as mentioned above, we notice that the distributions of both $\tau_{\rm x}$ and $\tau_{\rm p}$ develop a long non-exponential tail on the long-time end as $T$ diminishes (note that the logarithmic time scale is employed in Fig.~\ref{fig:decoupl}).
This is another indicator that the excitations at low $T$ do not obey Poisson statistics and thus are correlated.

\subsection{Analysis of threshold distance dependence}

In the previous subsection, we chose $d=3.0\,{\textrm \AA}$ as the threshold distance in the definition of excitations. 
The decoupling of the exchange and persistence times is present in a glassy environment with a physically meaningful value of $d$ chosen, because it is attributed to strong correlations of local events.
Here we examine how a different choice for the $d$ value would influence our analysis.  
For example, if we choose a $d$ value considerably less than the distance between neighboring ions, which is approximately $5\,{\textrm \AA}$ for our system (Fig.~\ref{fig:str}), excitations will correspond to small fluctuations of a central ion inside its counterion cage.  We can easily imagine that such excitations seldom induce a considerable structural change in the local environment. 
Excitations exceeding $\sim5.0\,{\textrm \AA}$ on the other hand describe delocalized hopping or gradual drift, which is usually accompanied by irreversible structural changes.  Thus, too large or too small a value for $d$ in the working definition of excitations would not be able to capture, e.g., the decoupling of persistence and the exchange times even though the decoupling occurs irrespective of our choice of $d$.

\begin{figure}
\centering
\resizebox{\textwidth}{!}{
	\includegraphics{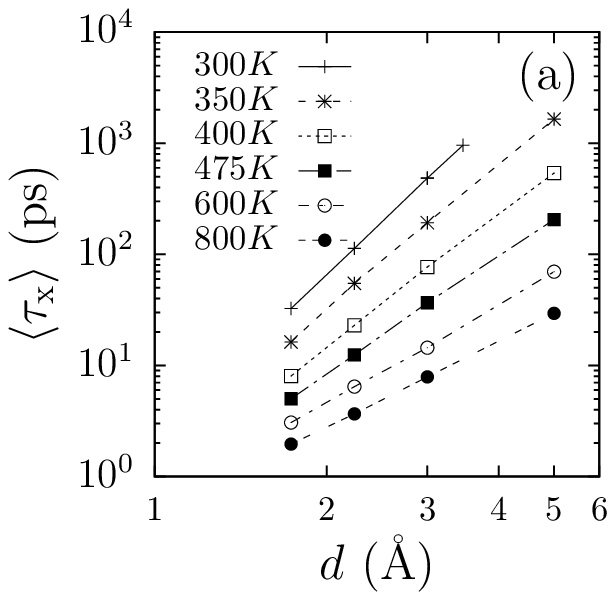}
	\includegraphics{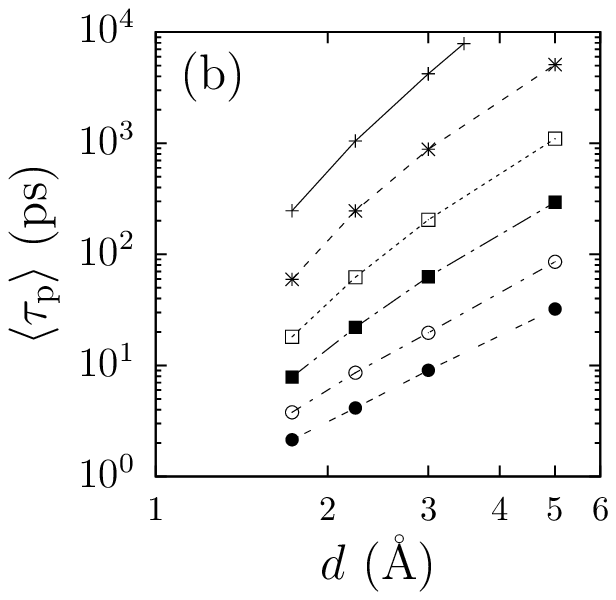}
}
\caption{Log-log plot of (a) the average exchange time $\langle \tau_{\mathrm x} \rangle$ and (b) the average persistence time $\langle \tau_{\mathrm p} \rangle$ for cations versus the threshold distance $d$. Lines are drawn as a guide for the eyes in (a) and (b).}
\label{fig:d}
\end{figure}

In Fig.~\ref{fig:d}, we display $\langle\tau_{\rm x}\rangle$ and 
$\langle\tau_{\rm p}\rangle$ of the cation versus $d$ at various temperatures. 
We notice that $\langle\tau_{\rm x}\rangle$ and $\langle\tau_{\rm p}\rangle$ increase with $d$ according to the power law, $\langle\tau_{\rm x,p}\rangle\sim d^{\delta}$, where $\delta$ ranges approximately from 2 at $800\,{\rm K}$ to 4 at $300\,{\rm K}$.  
In the scaling relation, the value $\delta\simeq2$ at $800\,{\rm K}$ corresponds to the  diffusive regime, whereas larger values at lower temperatures indicate anomalous diffusion.  Also the $d$-dependence of $\langle\tau_{\rm x}\rangle$ and $\langle\tau_{\rm p}\rangle$ becomes stronger as $T$ decreases.


\begin{figure}
\centering
\resizebox{\textwidth}{!}{
	\includegraphics{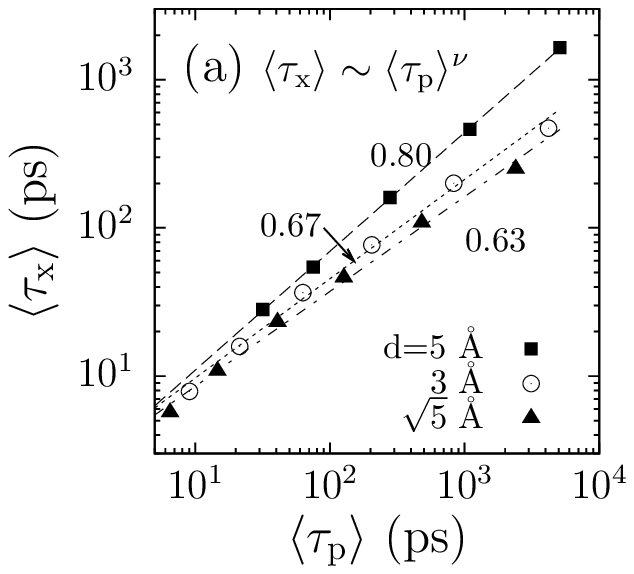}
	\hspace*{-20pt}
	\includegraphics{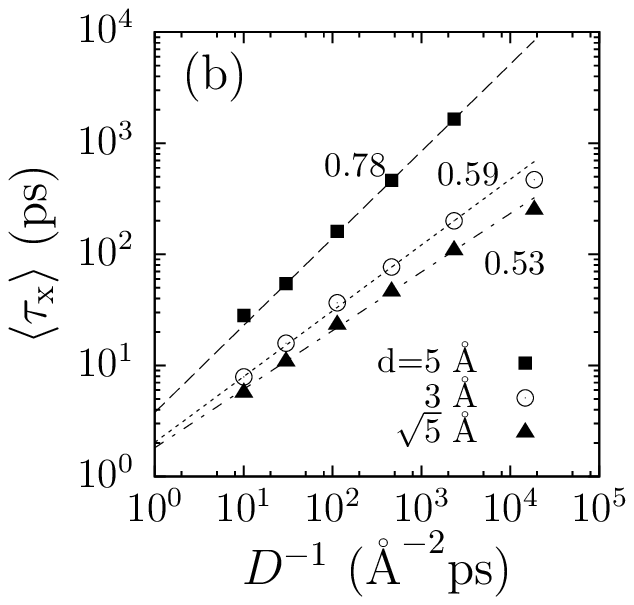}
	\hspace*{-20pt}
	\includegraphics{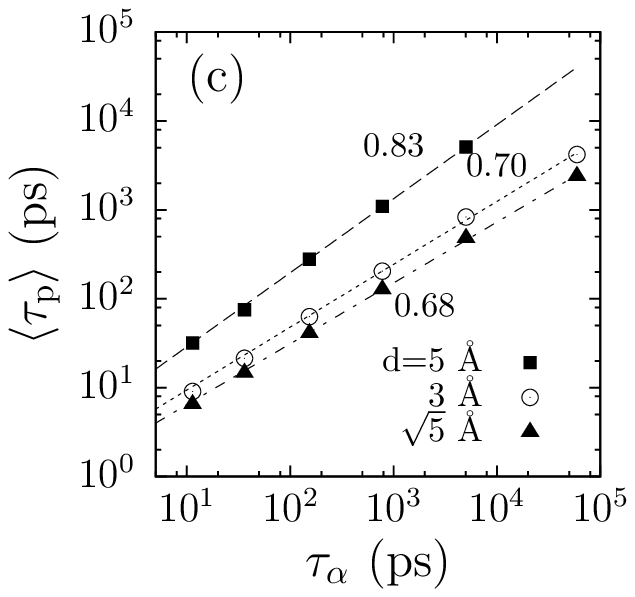}
}
\caption{(a) Average exchange time $\langle\tau_{\mathrm x}\rangle$ vs. average persistence time $\langle\tau_{\mathrm p}\rangle$; (b)  $\langle\tau_{\mathrm x}\rangle$ vs. inverse diffusion constant $D^{-1}$; (c) $\langle\tau_{\mathrm p}\rangle$ vs. structural relaxation time $\tau_{\alpha}$ for $d=5.0\,{\textrm \AA}$, $3.0\,{\textrm \AA}$, and $\sqrt{5}\,{\textrm \AA}$. 
Lines represent results fitted to (a) Eq.~\ref{eq:tauxtaup}; (b) and (c) the scaling behaviors $\sim x^{\gamma}$, where $x$ corresponds to $D^{-1}$ and  $\tau_{\alpha}$, respectively. In all cases, the value of the scaling exponents $\nu$ and $\gamma$, given for each $d$, is shown to increase with $d$.  
}
\label{fig:ddep}
\end{figure}

In Sec.~\ref{sec:SE}, the violation of the SE relation has been demonstrated via Eq.~\ref{eq:Dxi} 
with $\xi=0.87$ for cations. 
Likewise, the scaling relation 
\begin{equation}
\label{eq:tauxtaup}
\langle\tau_{\rm x}\rangle\sim\langle\tau_{\rm p}\rangle^\nu,
\end{equation}
is characterized by the exponent $\nu$, which is also smaller than unity. 
As shown in Fig.~\ref{fig:ddep}(a), with $d=5.0\,\textrm{\AA}$, $\nu$ is found to be $0.80$ for cations.  With $d=3.0$\,\AA, the corresponding $\nu$ value is
0.67.  
We point out that the decoupling between $\langle\tau_{\rm x}\rangle$ and $\langle\tau_{\rm p}\rangle$ appears greater than that of the SE violation, that is, $\nu$ is smaller than $\xi$ in Eq.~\ref{eq:Dxi}. It is known that $\tau_{\alpha}$  determined from $F_s(q_0, t)$ is identified with the average persistence time, while the exchange processes contribute to the diffusion if $d$ is comparable to the size of the ions ($d \gtrsim 5.0\,{\textrm \AA}$).~\cite{Chandler:prl,jung:exc,berthier:Fick_length}
In this perspective, we analyzed the scaling behaviors of $\langle \tau_{\rm x}\rangle$ with $D^{-1}$ and $\langle \tau_{\rm p}\rangle$ with $\tau_{\alpha}$. 
The results in Fig.~\ref{fig:ddep}(b) and (c) show that the former is characterized by weaker variations than the latter.
We also notice that the coupling behavior becomes stronger with $d$, i.e., the exponent in the scaling relation increases, for both cases.
But regardless of $d$, $\langle\tau_{\rm x}\rangle$ shows a significant sublinear behavior in $D^{-1}$. 
This is due to the fragile characteristic of our model system, where the exchange events are correlated.~\cite{jung:exc}
Therefore we expect that its scaling exponent would not reach unity even if a significantly larger value for $d$ is employed in the definition of $\langle\tau_{\rm x}\rangle$. 
By contrast, proportionality between $\langle\tau_{\rm p}\rangle$ and $\tau_{\alpha}$ is expected in the limit $d \rightarrow 2\pi/q_0$.
Note that the largest value of $d$ employed in Fig.~\ref{fig:ddep}(c) is $5.0\,{\textrm \AA}$, which is still less than $2\pi/q_0 (=7.32$\,\AA).  

\subsection{Comparison between excitation and brachiation}
\label{sec:brach}
\begin{figure}
\centering
\resizebox{\textwidth}{!}{
	\includegraphics{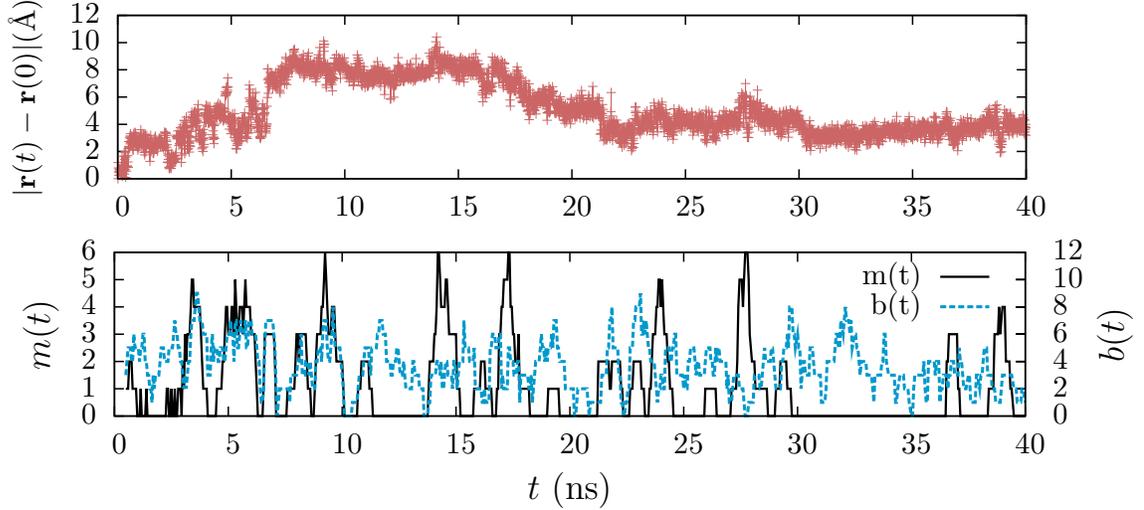}
}
\caption{Displacements of a cation and associated numbers of local excitation and brachiation events, $m(t)$ and $b(t)$.   $m(t)$ and $b(t)$ were determined by counting the excitations and brachiation events for the central cation occurring during a time window of $500\,{\rm ps}$ along its MD trajectory.  If neighboring anions revert back to original configurations within $200\,{\rm ps}$ of their initial changes, the corresponding events were not considered as brachiation according to Ref.~\citenum{Wu:brach}.
}
\label{fig:brach}
\end{figure}
In an effort to understand ion displacements and related diffusion in RTILs, a mechanism based on brachiation processes, viz., a central ions moves by forming and breaking links to neighboring counterions through Coulomb interactions, has been proposed recently.~\cite{Wu:brach}  While this has some similarity to local excitations and cage dynamics, it is essentially a structure-based description that lacks dynamic correlation effects.  To see this, we have performed a comparative analysis of brachiation and local excitations. In view of the brachiation length scale,\cite{Wu:brach} we reduced the cutoff distance $d$ slightly to $\sqrt{5}\,{\textrm \AA}$.  In Fig.~\ref{fig:brach}, the MD results for a typical cation at $350\,{\rm K}$ together with the numbers of local excitations and brachiation events, $m(t)$ and $b(t)$, are displayed.    We notice that while $m(t)$ and $b(t)$ show a significant overlap in certain part of the trajectory, there is in general no strong correlation between the two.  For instance, periods like $11 \,{\rm ns} < t< 13$\,ns and 
$30\, {\rm ns} < t< 36$\,ns are characterized by large $b(t)$ but vanishing $m(t)$. This means that neighboring ions reorganize in the presence of an immobile central ion.   The opposite situation where $m(t)$ is larger than $b(t)$, e.g., $23 \,{\rm ns} < t< 25$\,ns and
$27 \, {\rm ns} < t< 29$\,ns, also occurs frequently, indicating ion translations without any considerable change of its neighbors.  The weak correlation between local excitations and brachiation seems to suggest that the Coulomb interaction with neighboring ions does not play a major role in translational dynamics of individual ions in RTILs (see below).

\section{Role of Coulomb interaction}
\label{sec:coul}
Finally, we consider roles played by Coulomb interactions in the glassy behaviors of RTILs. \cite{shim:rtil:ver} 
To this end, we compare with a model supercooled liquid that has dynamic characteristics similar to our EMI$^+$PF$_6^-$ system but does not have long-range Coulombic interactions.


We employ a model liquid system with the WCA potentials of Ref.~\citenum{Hedges:decoupl} as a reference to quantify the influence of Coulomb interactions.  For easy comparison,  we follow Ref.~\citenum{Hedges:decoupl} to measure time and temperature in units of ${(m_{i}\sigma_{ii}^2/\epsilon_{ii})}^{1/2}$ and $\epsilon_{ii}/k_{\rm B}$, respectively.  We use the anion parameters, i.e., $i=$ PF$_6^-$, so that ${(m_{i}\sigma_{ii}^2/\epsilon_{ii})}^{1/2}=27.3$ ps and $\epsilon_{ii}/k_{\rm B}=180\,{\rm  K}$.  Thus $\tau_\alpha=5030\,{\rm ps}$ at $T=350\,{\rm K}$  corresponds to $\tilde\tau_{\alpha}=184$ at $\tilde T=1.94$ in the new scaled units.  The reader is reminded that the RTIL system at this temperature is characterized by strong nonexponential relaxation (Fig.~\ref{fig:ISF}).  By contrast, the WCA mixture does not exhibit 
glassy dynamics at all at this temperature.  In fact, dynamics comparable to our RTIL at $\tilde T=1.94$ occur at a much lower temperature $\tilde T=0.4$ for the WCA system.\cite{Hedges:decoupl}  This reveals that at a given (scaled) temperature, structural relaxation dynamics in EMI$^+$PF$_6^-$ in the CGM description are considerably slower than those in supercooled liquids characterized by short-range interactions only.  This indicates that the Coulomb interactions in effect suppress structural fluctuations and enhance trapping in cage structures.

To gain insight at the microscopic level, we briefly analyze instantaneous power inputs from the RTIL environment to an ion $i$
via the Coulomb and LJ forces, i.e., ${\bf F}^{\rm Coul}_i\cdot {\bf v}_i$ and ${\bf F}^{\rm LJ}_i\cdot {\bf v}_i$ (${\bf v}_i=$ ion velocity). 
For simplicity, we consider only the united atom T1 for cations because it is the heaviest atom, located close to the cation center of mass.  We examined the static correlation between the two instantaneous powers using
the Pearson correlation coefficient $\rho_{X,Y}$:~\cite{Kanji}
\begin{equation}
\rho_{X,Y}=\frac{\langle (X-\langle X\rangle)(Y-\langle Y \rangle)\rangle}
                {\sigma_X \sigma_Y},
\end{equation}
where $X$ and $Y$ denote ${\bf F}^{\rm Coul}_i\cdot {\bf v}_i$ and ${\bf F}^{\rm LJ}_i\cdot {\bf v}_i$, respectively, and $\sigma_X$ and $\sigma_Y$ are their standard deviations. 
\begin{figure}
\centering
\resizebox{0.5\textwidth}{!}{
	\includegraphics{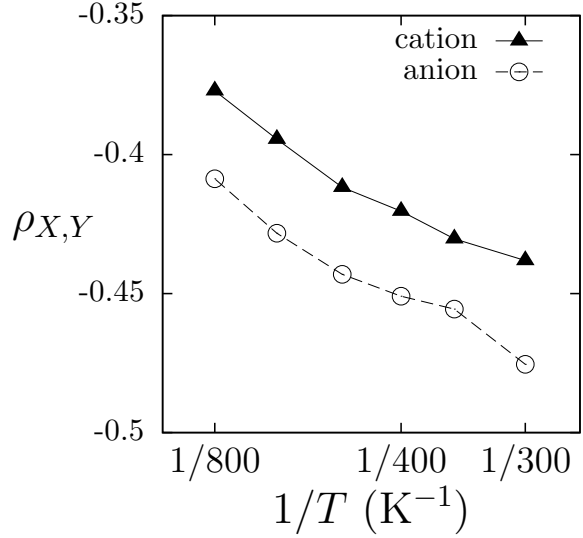}
}
\caption{Correlation coefficient $\rho_{X,Y}$ versus the inverse temperature, where $X$ and $Y$ denotes 
the instantaneous powers by the Coulomb force and the Lennard-Jones force. 
As the temperature is lowered, the negativity of linear correlations between $X$ and $Y$ becomes apparent.
}
\label{fig:cc}
\end{figure}
The results in Fig.~\ref{fig:cc} disclose that power inputs arising from the LJ and Coulomb forces are anti-correlated.  
Therefore, LJ and Coulomb interactions play antagonistic roles in energy relaxation of individual ions.  If we calculate the time integration of ${\bf F}^{\rm Coul}_i\cdot {\bf v}_i$ and ${\bf F}^{\rm LJ}_i\cdot {\bf v}_i$, however, the contribution from the former nearly vanishes and work to individual ions is delivered primarily by the LJ force(data not shown).  This result together with our comparative analysis above paints the picture that while the liquid structure of RTILs and thus the energy scale, i.e., $\tilde T$, relevant for glassy dynamics are mainly determined by Coulomb interactions, their relaxation is generally governed by the LJ interactions.  We believe this explains why dynamic aspects of glassy RTILs are very similar to those of non-ionic supercooled liquids despite their major difference in long-range interactions. We note that the anti-correlation of power inputs from the LJ and Coulomb forces and the dominance of the former in relaxation dynamics of the ionic liquid observed here also apply to vibrational energy relaxation in RTILs.\cite{shim:rtil:ver}



\section{Conclusions}
\label{sec:conclusion}
We have introduced the coarse-grained model of ionic liquids and probed its dynamical behaviors.
Coarse-graining has simplified the geometry of the system and made dynamics accelerated, compared with the atomistic model.  Nevertheless, the overall liquid structure and glassy dynamic properties such as nonexponential structural relaxation and subdiffusive behavior are preserved.  
Owing to the reduced number of atoms by coarse-graining, we have been able to perform extensive MD simulations at long times over a wide range of temperature, and investigated the temperature dependence of structural relaxation and diffusion. 

We found that our model for ionic liquids belongs to fragile glass formers, where $\tau_{\alpha}$ exhibits strong non-Arrhenius dependence on the temperature.  
In addition, the SE relation is violated, which implies that diffusion is enhanced when compared with structural relaxation. 
We pay attention to the apparent universality of the abovementioned dynamic features observed in a variety of glassy liquid systems, regardless of the nature of their molecular interactions. 
In previous studies of supercooled liquids, kinetic constraints have been successfully employed to explain the peculiar dynamic properties of supercooled liquids in view of facilitated dynamics. Dynamic facilitation emphasizes the dominant role of dynamic correlations in glassy dynamics rather than static properties such as the structure factor and potential energy landscape.~\cite{Stillinger:landscape}
In a similar fashion, we have defined a local excitation in the dynamics of our model and observe dynamic correlations between them.  Decoupling of persistence and exchange times has been shown to be highly correlated with local excitations. 
Such decoupling behavior exists regardless of the threshold distance $d$, which defines local excitations. However, $d$ determines the physical meaning of local excitation events, especially at low temperatures. 

To understand the influence of the Coulomb interactions, we compared structural relaxation dynamics of the RTIL with those of non-ionic models of supercooled liquids.  We found that glassy dynamics occur at a considerably higher temperature (in scaled dimensionless units) for the former than for the latter.
We have also investigated instantaneous powers arising from the Coulomb and LJ forces.  It was found that they are anti-correlated and their time integration is dominated by the latter.  These results seem to indicate that relaxation dynamics of RTILs are dominated by the LJ interactions, while the Coulomb interactions exert a strong influence on the liquid structure and thus set the temperature scale for glassy dynamics. 


At low temperatures, immobile regions persist for a long time due to sparse excitations. Once a local excitation occurs, subsequent displacements of ions are more probable, which tends to introduce mobile regions. Thus, decoupling of persistence and exchange times is a plausible explanation for dynamic heterogeneity of glassy liquids.  
In the future, we plan to investigate spatiotemporal  correlations of local excitation events in order to characterize in more detail dynamic heterogeneity in the RTIL. 

\section*{Acknowledgments}
This work was supported by the National Research Foundation of Korea (Grant Nos. R11-2007-012-03003-0, 2009-0070517, and 2009-0080791), the KISTI Supercomputing Center (KSC-2007-500-3008 and KSC-2009-502-0003), and the BK21 Program. 

\bibliographystyle{rsc}
\bibliography{ref_pccp_HK}

\end{document}